\documentclass[runningheads]{llncs}

\usepackage[T1]{fontenc}
\usepackage{graphicx}
\usepackage{subfig}
\usepackage{hyperref}

\begin{document}
\title{Interpretable Weighted Siamese Network to Predict the Time to Onset of Alzheimer's Disease from MRI Images}


\titlerunning{Interpretable Weighted Siamese Network}
%

\author{Misgina Tsighe Hagos\inst{1,2}\orcidID{0000-0002-9318-9417} \and
Niamh Belton\inst{1,3}\orcidID{0000-0003-4949-4745} \and
Ronan P. Killeen\inst{3,4,5}\orcidID{0000-0002-9231-8133} \and
Kathleen M. Curran\inst{1,3}\orcidID{0000-0003-0095-9337}\and
Brian Mac Namee\inst{1,2}\orcidID{0000-0003-2518-0274} \and for the Alzheimer's Disease Neuroimaging Initiative\thanks{Data used in preparation of this article were obtained from the Alzheimer’s Disease Neuroimaging Initiative (ADNI) database (adni.loni.usc.edu). As such, the investigators within the ADNI contributed to the design and implementation of ADNI and/or provided data but did not participate in analysis or writing of this report. A complete listing of ADNI investigators can be found at: \url{https://adni.loni.usc.edu/wp-content/uploads/how\_to\_apply/ADNI\_Acknowledgement\_List.pdf}}}

\authorrunning{M. T Hagos et al.}
%
\institute{Science Foundation Ireland Centre for Research Training in Machine Learning \email{misgina.hagos@ucdconnect.ie} \hspace{0.25cm} \url{http://www.ml-labs.ie}\and
\mbox{School of Computer Science, \and School of Medicine,} University College Dublin\\
\and
\mbox{Department of Radiology, \and UCD–SVUH PET CT Research Centre, }\\St Vincent’s University Hospital, Dublin 4, Ireland}



%
\maketitle              
\begin{abstract}
Alzheimer's Disease (AD) is a progressive disease preceded by Mild Cognitive Impairment (MCI). Early detection of AD is crucial for making treatment decisions. However, most of the literature on computer-assisted detection of AD focuses on classifying brain images into one of three major categories: healthy, MCI, and AD; or categorizing MCI patients into (1) \emph{progressive}: those who progress from MCI to AD at a future examination time, and (2) \emph{stable}: those who stay as MCI and never progress to AD. This misses the opportunity to accurately identify the trajectory of progressive MCI patients. In this paper, we revisit the brain image classification task for AD identification and re-frame it as an ordinal classification task to predict \emph{how close a patient is to the severe AD stage.} To this end, we select progressive MCI patients from the Alzheimer's Disease Neuroimaging Initiative (ADNI) dataset and construct an ordinal dataset with a prediction target that indicates the time to progression to AD. We train a Siamese network model to predict the time to onset of AD based on MRI brain images. We also propose a Weighted variety of Siamese network and compare its performance to a baseline model. Our evaluations show that incorporating a weighting factor to Siamese networks brings considerable performance gain at predicting how close input brain MRI images are to progressing to AD. Moreover, we complement our results with an interpretation of the learned embedding space of the Siamese networks using a model explainability technique.

\end{abstract}
\section{Introduction}
\label{section:introduction}

Although it has been more than a century since Alois Alzheimer first described the clinical characteristics of Alzheimer's Disease (AD) \cite{alzheimer1907uber,alzheimer1995english}, the disease still eludes early detection. AD is the leading cause of dementia, accounting for 60\%-80\% of all cases worldwide \cite{kalaria2008alzheimer,zhang2017advancing}, and the number of patients effected is growing. In 2015, Alzheimer's Disease International (ADI) reported that over 46 million people were estimated to have dementia worldwide and that this number was expected to increase to 131.5 million by 2050 \cite{prince2015world}. Since AD is a progressive disease, computer assisted early identification of the disease may enable early medical treatment to slow its progression.

Methods that require intensive expert input for feature collection, such as Morphometry \cite{dashjamts2012alzheimer}, and more automated solutions based on deep learning \cite{billones2016demnet,liu2018multi,basaia2019automated} have been utilized in the computer assisted diagnosis of AD literature. These automated detection methods usually classify patients as belonging to one of three stages: Normal (patients exhibiting no signs of dementia and no memory complaints), Mild Cognitive Impairment (MCI) (an intermediate state in which a patient's cognitive decline is greater than expected for their age, but does not interfere with activities of their daily life), and full AD. 

\begin{figure}[h]
    \centering
    \subfloat[\centering Progression levels]{{\includegraphics[width=5.5cm]{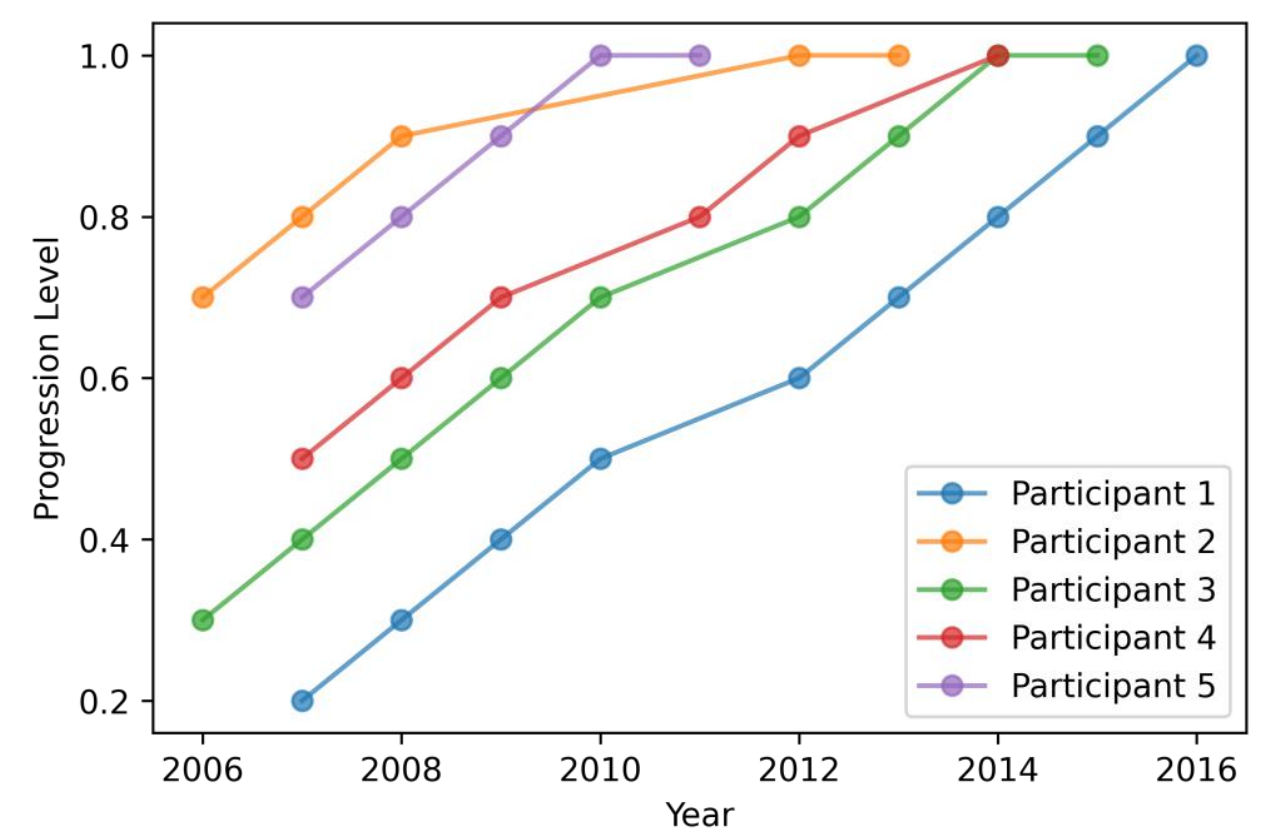} }}%
    \qquad
    \subfloat[\centering Traditional data preparation]{{\includegraphics[width=5.5cm]{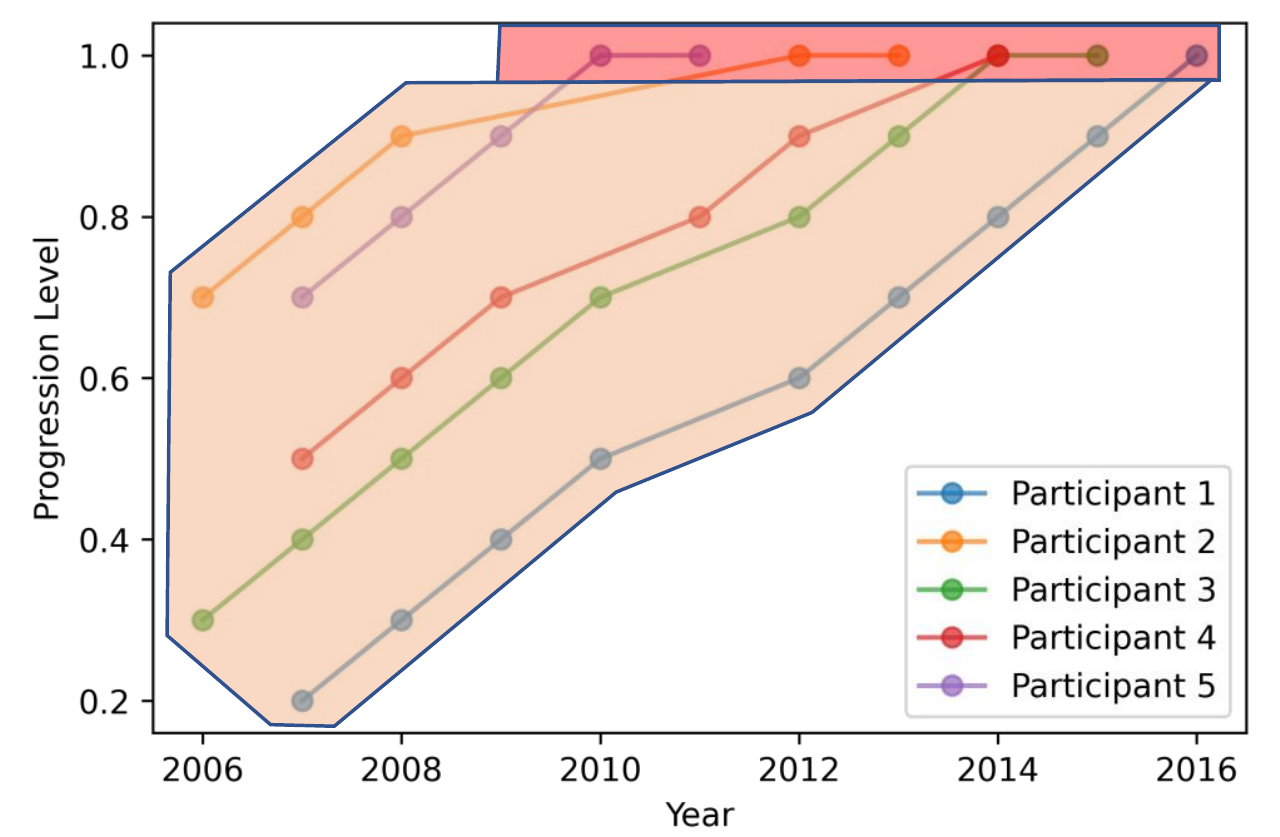} }}%
    \qquad
    \subfloat[\centering Our approach]{{\includegraphics[width=6cm]{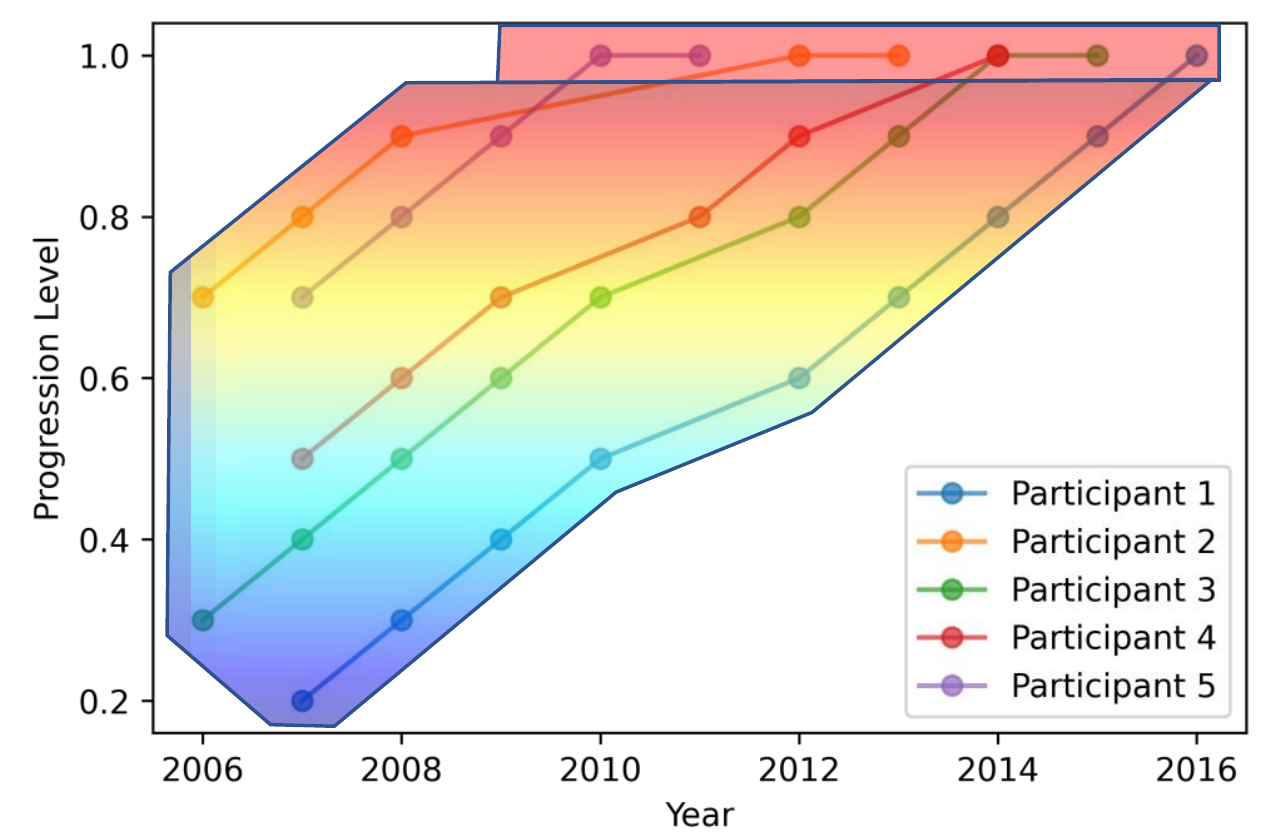} }}%
    \caption{(a) Progression levels of five sample MCI participants where each dot represents an MRI image during an examination year. (b) shows the typical approach of organizing images for identification of progressive MCI or classification of MCI and AD. The images in the lower light orange section are categorized as MCI when preparing a training dataset (this includes those images from patients nearing progression to AD---near progression level 1.0). The images in the upper red section are categorized as AD. In (c) our approach to organizing brain images is illustrated using a Viridis color map. Images are assigned ordinal progression levels $\in[ 0.1,0.9 ]$ based on their distance in years from progressing to AD stage.} 
    \label{figure:progression_levels}%
\end{figure}

A participant's progression from one of the stages to the next, however, can take more than five years \cite{roberts2014higher}. This can mean that when automated disease classification systems based on these three levels are used, patients at a near severe stage do not receive the required treatment because they are classified as belonging to the pre-severe stage. This is illustrated in Fig. \ref{figure:progression_levels}(a) for five participants from the Alzheimer's Disease Neuroimaging Initiative (ADNI) study \cite{mueller2005ways}. Using the typical classification approach (See Fig. \ref{figure:progression_levels}(b)), for example, even though participant five (purple) is only a year away from progressing to the severe AD stage, they would be classified as MCI in 2009. To address this issue we focus on the clinical question ``how far away is a progressive MCI patient on their trajectory to AD?'' To do this we propose an ordinal categorization of brain images based on participants' level of progression from MCI to AD as shown in Fig \ref{figure:progression_levels}(c). Our approach adds ordinal labels to MRI scans of patients with progressive MCI indicating how many years they are from progressing to AD, and we construct a dataset of 444 MRI scans from 288 participants with these labels and share a replication script.


In addition to constructing the dataset, we also develop a computer assisted approach to identifying a participant's (or more specifically, their MRI image's) progression level. Accurately identifying how far a patient is from progressing to full AD is of paramount importance as this information may enable earlier intervention with medical treatments \cite{albright2019forecasting}. Rather than using simple ordinal classification techniques, we use Siamese networks due to their ability to handle the class imbalance in the employed dataset \cite{koch2015siamese,yang2020hscnn}. We use a Siamese network architecture, and a novel Weighted Siamese network that uses a new loss function tailored to learning to predict input MRI image’s likelihood of progression. Furthermore, we complement results of our Siamese network based method with interpretations of the embedding space using an auxiliary model explanation technique, T-distributed Stochastic Neighbor Embedding (t-SNE) \cite{van2008visualizing}. t-SNE condenses high dimensional embedding spaces learned by a Siamese network into interpretable two or three dimensional spaces \cite{borys2023explainable}.




The main contributions of this paper are:

\begin{enumerate}
    \item We provide a novel approach that interpolates ordinal categories between existing MCI and AD categories of the ADNI dataset based on participants' progression levels.
    \item We apply the first Siamese network approach to predict interpolated progression levels of MCI patients.
    \item We propose a simple and novel variety of triplet loss for Siamese networks tailored to identifying progression levels of MCI patients.
    \item Our experiments demonstrate that using our version of the triplet loss is better at predicting progression level than the traditional triplet loss. Code is shared online\footnote{\url{https://github.com/Msgun/WeightedSiamese}}.
\end{enumerate}

\section{Related Work}

Before the emergence of deep learning, and in the absence of relevant large datasets, computer-assisted identification of AD relied on computations that require expensive expert involvement such as Morphometry \cite{dashjamts2012alzheimer}. However, the release of longitudinal datasets, such as ADNI \cite{mueller2005ways}, inspired research on automated solutions that employed machine learning and deep learning methods for the identification of AD. 

Most of the approaches proposed for AD diagnosis perform a classification among three recognized stages of the disease: Normal, MCI, and AD \cite{li2018classification}. Some examples in the literature distinguish between all three of the categories \cite{zhou2019effective}, while others distinguish between just two: Normal and MCI \cite{ju2019early}, Normal and AD \cite{xiao2021early}, or MCI and AD \cite{wegmayr2018classification}.

Patients at the MCI stage have an increased risk of progressing to AD, especially for elderly patients \cite{roberts2014higher}. For example, in the Canadian Cohort Study of Cognitive Impairment and Related Dementia \cite{hsiung2006outcomes} 49 out of a cohort of 146 MCI patients progressed to AD in a two-year follow-up. In general, while healthy adult controls progress to AD annually at a maximum rate of 2\%, MCI patients progress at a rate of 10\%-25\% \cite{grand2011clinical}. This necessitates research on identifying MCI subjects at risk of progressing to AD. In a longitudinal study period, participants diagnosed with MCI can be categorized into two categories: (1) Progressive MCI, which represents participants who were diagnosed with MCI at some stage during the study but were later diagnosed with AD, and (2) Stable MCI, patients who stayed as MCI during the whole study period \cite{hagos2022interpretable}. This excludes MCI participants with chances of reverting back to healthy, since they were also reported to have chances of progressing to AD \cite{roberts2014higher}. There are some examples in the literature of using machine learning techniques such as random forest \cite{moradi2015machine} and CNNs \cite{hagos2022interpretable} to classify between stable and progressive MCI. While feature extraction is used prior to model training towards building relatively simpler models \cite{li2018classification,zheng2018early}, 3D brain images are also deployed with 3D CNNs to reduce false positives \cite{basaia2019automated,liu2018anatomical}.


The brain image classification task can also be transformed to ordinal classification to build regressor models. For example, four categories of AD: healthy, stable MCI, progressive MCI, and AD were used as ordinal labels to build a multi-variate ordinal regressor using MRI images in \cite{doyle2014predicting}. However, the output of these models gives no indication of the likelihood of a patient to progress from one stage to another. Furthermore, this does not provide prediction for interpolated inter-category progression levels. Albright et al. (2019) \cite{albright2019forecasting} used a longitudinal clinical data including ADAS13, which is a 13-item Alzheimer’s Disease Assessment Scale, and Mini-Mental State Examination (MMSE) to train multi-layer perceptron and recurrent networks for AD progression prediction. This work, however, uses no imaging data and it has been shown that brain images play a key role in improving diagnostic accuracy for Alzheimer's disease \cite{johnson2012brain}. 

Siamese networks, which use a distance-based similarity training approach \cite{bromley1993signature,chopra2005learning}, have found applications in areas such as object tracking \cite{bertinetto2016fully} and anomaly detection \cite{belton2021semi}. Although it does not focus on AD detection, we found \cite{li2020siamese} to be the closest approach to our proposed method in the literature. Li et al. \cite{li2020siamese} report that a Siamese network's distance output could be translated to predict disease positions on a severity scale. Although this approach takes the output as severity scale without any prior training on disease severity, only deals with existing disease stages, and does not interpolate ordinal categories within, it does suggest Siamese networks as a promising approach for predicting ordinal progression levels.

\section{Approach}
\label{section:experiments}

In this section, we describe the datasets (and how they are processed), model architectures, model training and evaluation techniques used in our experiments, as well as our proposed triplet loss for Siamese networks.

\subsection{Dataset Preparation}

The data used in the experiments described here was obtained from the Alzheimer's Disease Neuroimaging Initiative (ADNI) database. ADNI was launched in 2003, led by Principal Investigator Michael W. Weiner, MD (\url{adni.loni.usc.edu}). For up-to-date information, see \url{www.adni-info.org}. 


\begin{table}
\centering
\caption{Image distribution across progression levels, $\rho$.}
\label{table:image_distribution}
\begin{tabular}{l r r r r r r r r r}
\hline
$\rho$                & 0.2 & 0.3 & 0.4 & 0.5 & 0.6 & 0.7 & 0.8 & 0.9 & 1.0 \\
\hline
Number of images & 4   & 4   & 6   & 10  & 24  & 56  & 172 & 273 & 467\\
\hline
\end{tabular} 
\end{table}

\noindent From the ADNI dataset we identified MRI brain images of 1310 participants who were diagnosed with MCI or AD. 288 participants had progressive MCI, 545 had stable MCI, and the rest had AD. We used MRI images of the 288 progressive MCI participants to train and evaluate our models. We labeled the progressive MCI participants based on their progression levels towards AD, $\rho \in [0.1, 1.0]$ with a step size = 0.1, where for a single participant, $P$, $min(\rho)=0.1$ represents the first time $P$ was diagnosed with MCI and $P$ transitions to stage AD at $max(\rho)=1.0$. This transforms the binary MCI and AD labels to 10 ordinal labels. An example of the data organization based on progression level is plotted in Fig. \ref{figure:progression_levels}. The distribution of the constructed ordinal regression levels is shown in Table \ref{table:image_distribution} (where $\rho=1.0$ represents AD) where the imbalance between the different labels is clear. Within the ADNI dataset, the maximum number of MRI scans that the progressive MCI participants have had until they progressed to AD ($\rho=1.0$) is 9, which means that the smallest $\rho$ is 0.2. We took advantage of Siamese networks robustness to class imbalance to circumnavigate the imbalance in the ordinal labels. By sub-sampling from the majority classes, we selected 444 3D MRI images (shape = 160x192x192) for the negative, anchor, and positive datasets (each holding 148 images) required when training a Siamese network using triplet loss. We used 80\% of the images for training and the rest for testing. AD images were randomly separated to the anchor and positive dataset. We ensure that there is no participant overlap between sets when performing the data splitting between training and testing dataset, and between anchor and positive dataset.

\subsection{Weighted Siamese Network}

\begin{figure}
    \centering
    \centering
    \includegraphics[width=1.\linewidth]{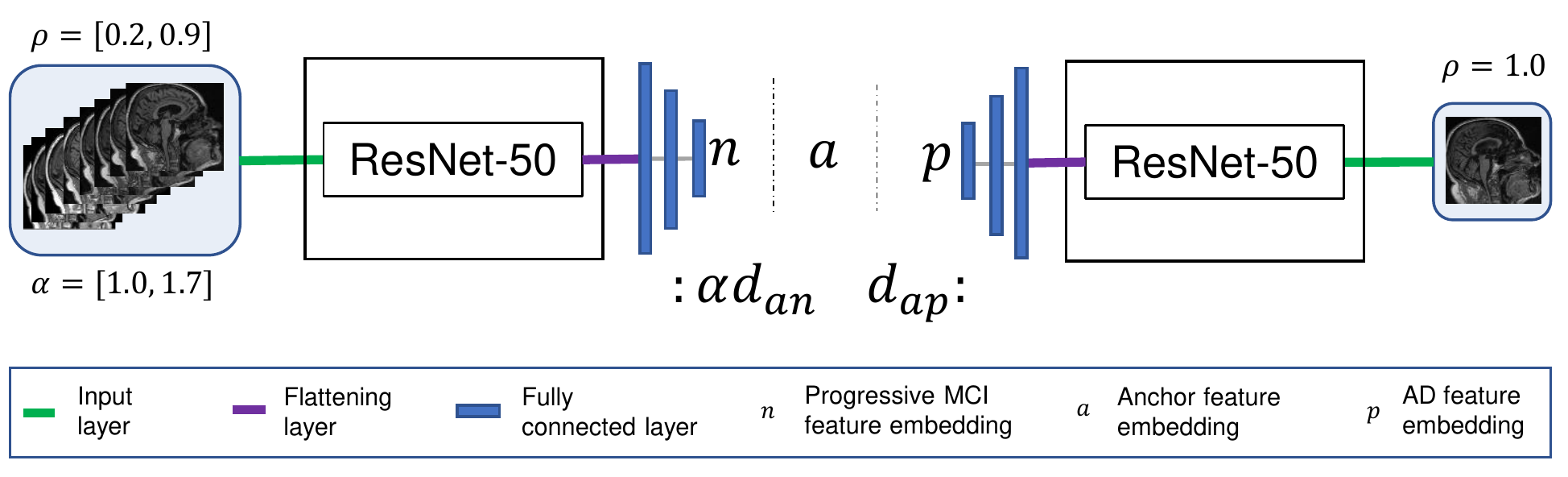}
    \caption{Weighted Siamese network. The text ResNet-50 here refers to the base layers of the ResNet-50 architecture which are trained from scratch for extracting image embeddings, i.e excluding the fully connected classifier layers. While we used 3D MRI images for model training and evaluation, sagittal plane is used here only for visualization purposes. (The figure is best viewed in color.)}%
    \label{figure:weighted_siamese_architecture}%
\end{figure}

\begin{figure}
    \centering
    \centering
    \includegraphics[width=7.5cm]{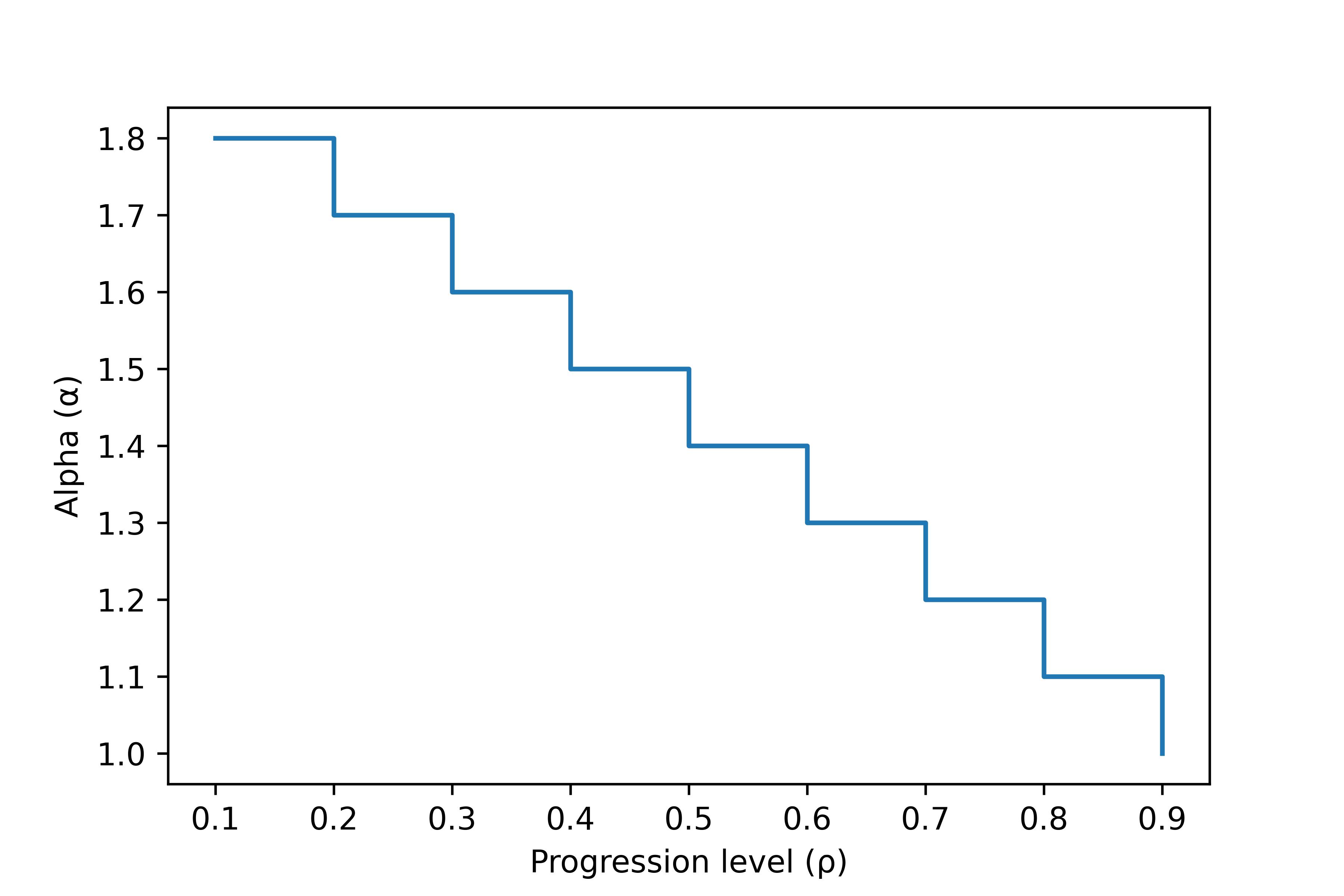}
    \caption{Transforming progression level, $\rho$, to $\alpha$.}%
    \label{figure:progression_level_vs_alpha}%
\end{figure}

\noindent Siamese networks are usually trained using a triplet loss or its variants. While a traditional triplet loss teaches a network that a negative instance is supposed to be at a larger distance from the anchor than a positive instance, we propose a Weighted triplet loss that teaches a network that instances, which can all be considered to be in the negative category, are not at the same distance from the anchor and that their distance depends on their progression level, $\rho$. So that lower progression levels have larger distance from an anchor instance, we transform $\rho$ to a weighting coefficient $\alpha = 1.9 - \rho$, excluding $\rho=1.0$, as shown in Fig. \ref{figure:progression_level_vs_alpha}. The architecture of our proposed Weighted Siamese network is shown in Fig. \ref{figure:weighted_siamese_architecture}.

\noindent We used two different loss functions to train our Siamese networks. The first is a traditional triplet loss, which we refer to as Unweighted Siamese: 

\begin{equation}
\label{equation:unweighted}
    L_{u} = max(d_{ap} - d_{an} + margin, 0)
\end{equation}

\noindent where  $margin=1.0$, $d_{ap}$ is the Euclidean distance between anchor and positive embeddings, and $d_{an}$ is the distance between the embeddings of anchor and negative instances.

The second loss is a newly proposed Weighted triplet loss---Weighted Siamese which introduces a coefficient $\alpha \in [1.0, 1.8]$ to $d_{an}$ in $L_{u}$:

\begin{equation}
\label{equation:weighted}
L_{w} = max(d_{ap} - \alpha  d_{an} + margin, 0)
\end{equation}

\subsection{Training and Evaluation}

We implemented all of our experiments using TensorFlow \cite{tensorflow2015-whitepaper} and Keras \cite{chollet2015keras}. After comparing performance between different architectures and feature embedding size, we chose to train a 3D ResNet-50 model from scratch by adding three fully connected layers of sizes 64, 32, and 8 nodes with ReLu activations, taking the last layer of size 8 as the embedding space. We used an Adam optimizer with a decaying learning rate of 1e-3. We trained the model with five different seeds for 150 epochs, which took an average of 122 minutes per a training run on an NVIDIA RTX A5000 graphics card.

For model evaluation on training and testing datasets, we use both the Unweighted Siamese and Weighted Siamese losses as well as Mean Absolute Error (MAE) and Root Mean Squared Error (RMSE). MAE and RMSE are presented in Equations \ref{equation:mae} and \ref{equation:rmse} respectively, for a test set of size $N$ where $y_{i}$ and $Y_{i}$ hold the predicted and ground truth values for instance $i$, respectively. We turn the distance outputs of the Siamese networks into $y$ by discretizing them into equally spaced bins, where the number of bins equals the number of progression levels. 

\begin{equation}
\label{equation:mae}
    MAE = \frac{1}{N} \sum_{i=1}^N| y_{i} - Y_{i} |
\end{equation}

\begin{equation}
\label{equation:rmse}
    RMSE = \sqrt{\frac{1}{N} \sum_{i=1}^N (y_{i} - Y_{i})^2}
\end{equation}

\noindent We make use of t-SNE to explain the 8 dimension embedding space learned by the Weighted Siamese network by condensing it two dimensions. For presentation purposes and in order to fit the t-SNE well, we drop underrepresented progression levels; while we dropped progression level 0.2 from the training dataset, progression levels 0.2, 0.3, and 0.5 were removed from the testing dataset. The t-SNE was fitted over a 1000 iterations using Euclidean distance metric with a perplexity of 32 and 8 for the training and testing datasets, respectively. 


\section{Results and Discussion}
\label{section:results}

In this section, we present training and testing losses, MAE and RMSE metrics of evaluation, a plot showing comparison between predicted and ground-truth progression levels, as well as interpretation of the results.

\begin{figure}[t]
    \centering
    \subfloat[\centering Training loss]{{\includegraphics[width=5.6cm]{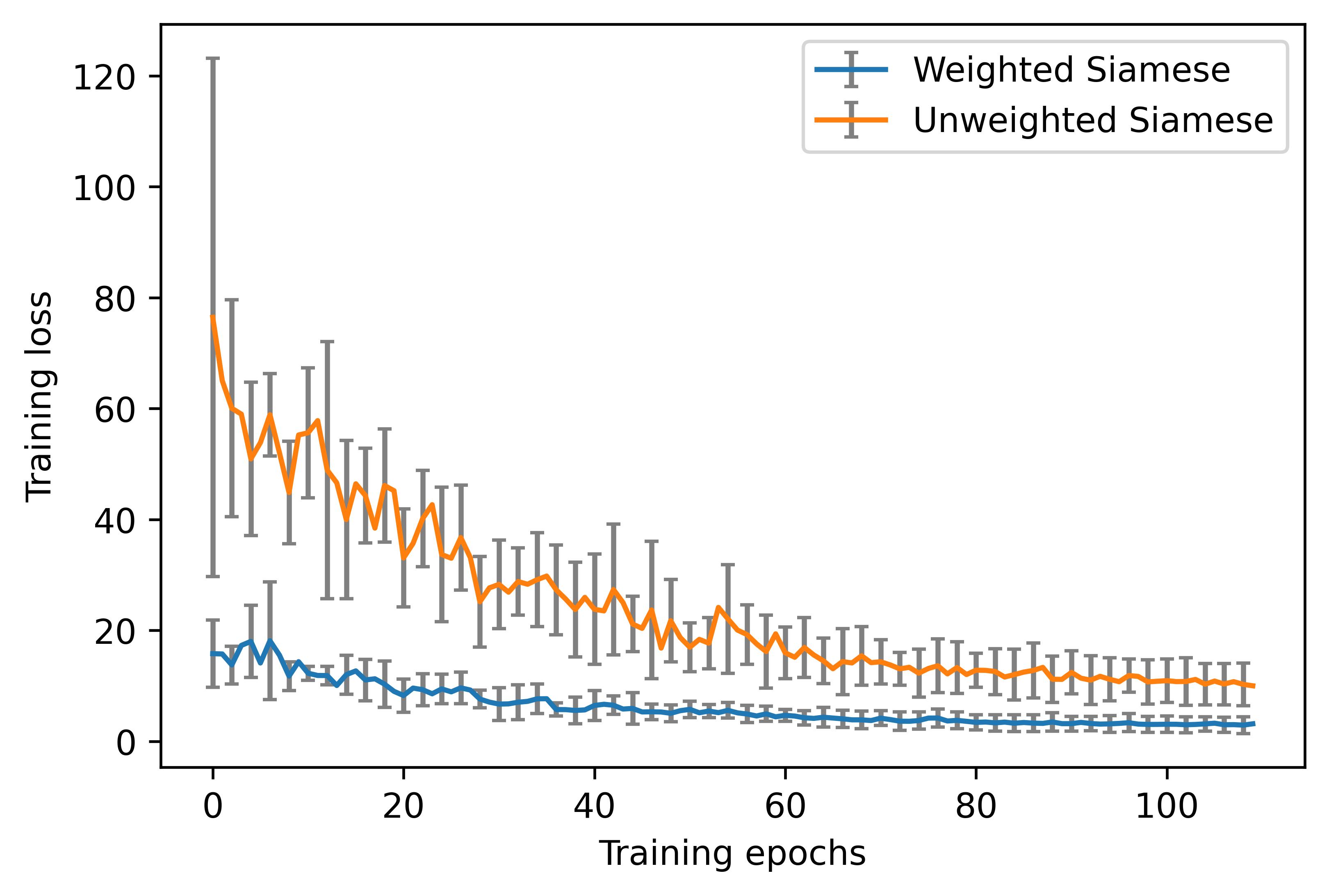} }}%
    \qquad
    \subfloat[\centering Test loss]{{\includegraphics[width=5.6cm]{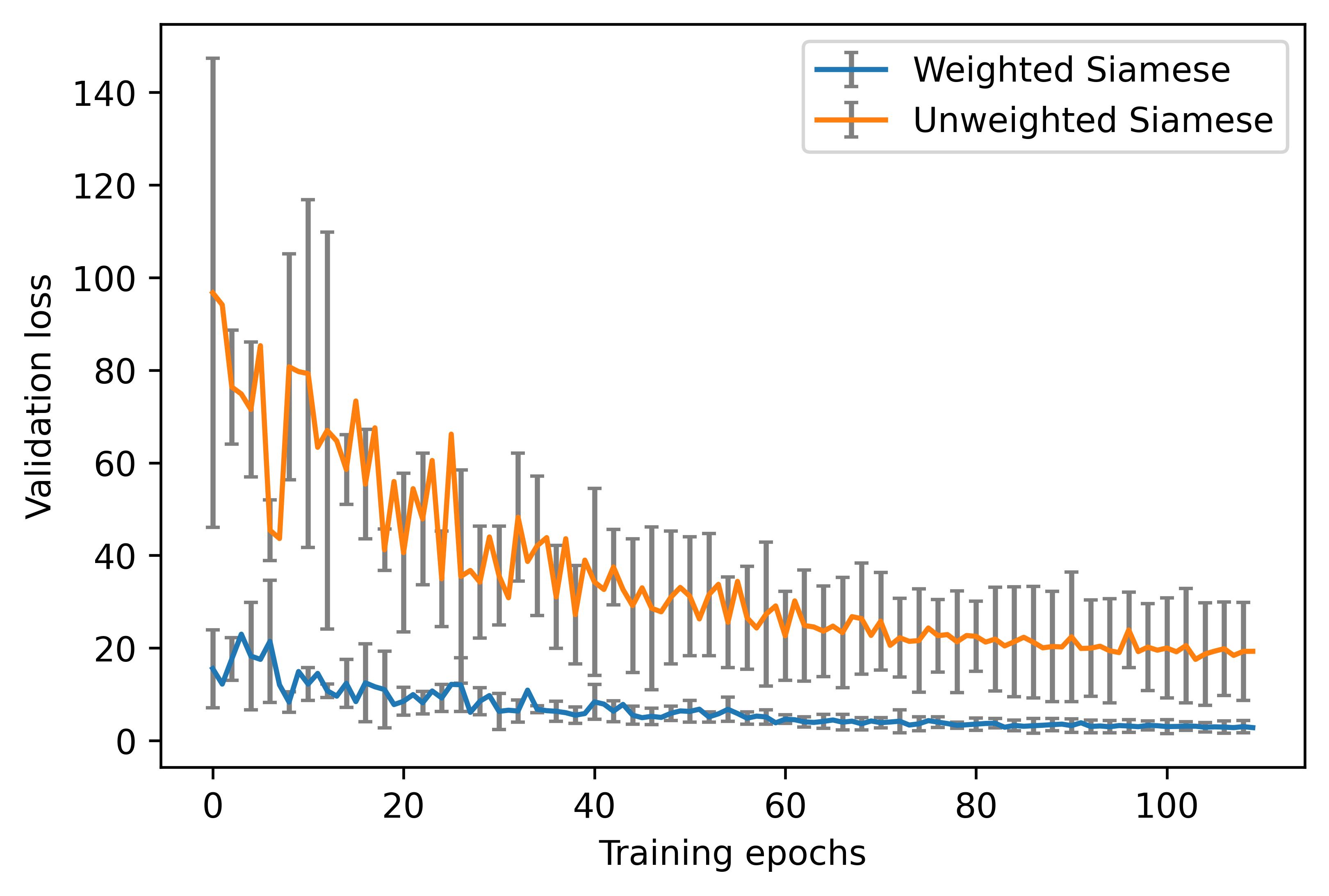} }}%
    
    \caption{Training and testing losses of the Weighted and Unweighted Siamese models. The first 40 epochs are cropped out for easier visualization. Bars represent std. errors over five runs.}%
    \label{figure:losses}%
\end{figure}


\begin{table}[b]
\centering
\caption{Average MAE and RMSE over five runs.}
\label{table:average_mae_rmse}
\begin{tabular}{lrr}
\hline
Method &  MAE & RMSE\\
\hline
Unweighted Siamese &  2.30 & 2.94 \\
Weighted Siamese &  2.00 & 2.40\\
\hline
\end{tabular}
\end{table}

Training and testing losses over five runs of model training for both the Unweighted and Weighted Siamese networks are shown in Fig.
\ref{figure:losses}. While the average training and testing losses of the Weighted Siamese network are 2.92 and 2.79, the Unweighted Siamese achieves 10.02 and 17.53, respectively. We were able to observe that the Unweighted Siamese network had  a hard time learning the progression levels of all the ordinal categories. However, our proposed approach using Weighted loss was better at fitting to all the levels. We accredit this to the effects of adding a weighing factor using $\rho$.

\begin{figure}[!t]
    \centering
    \subfloat[\centering Weighted Siamese]{{\includegraphics[width=5.6cm]{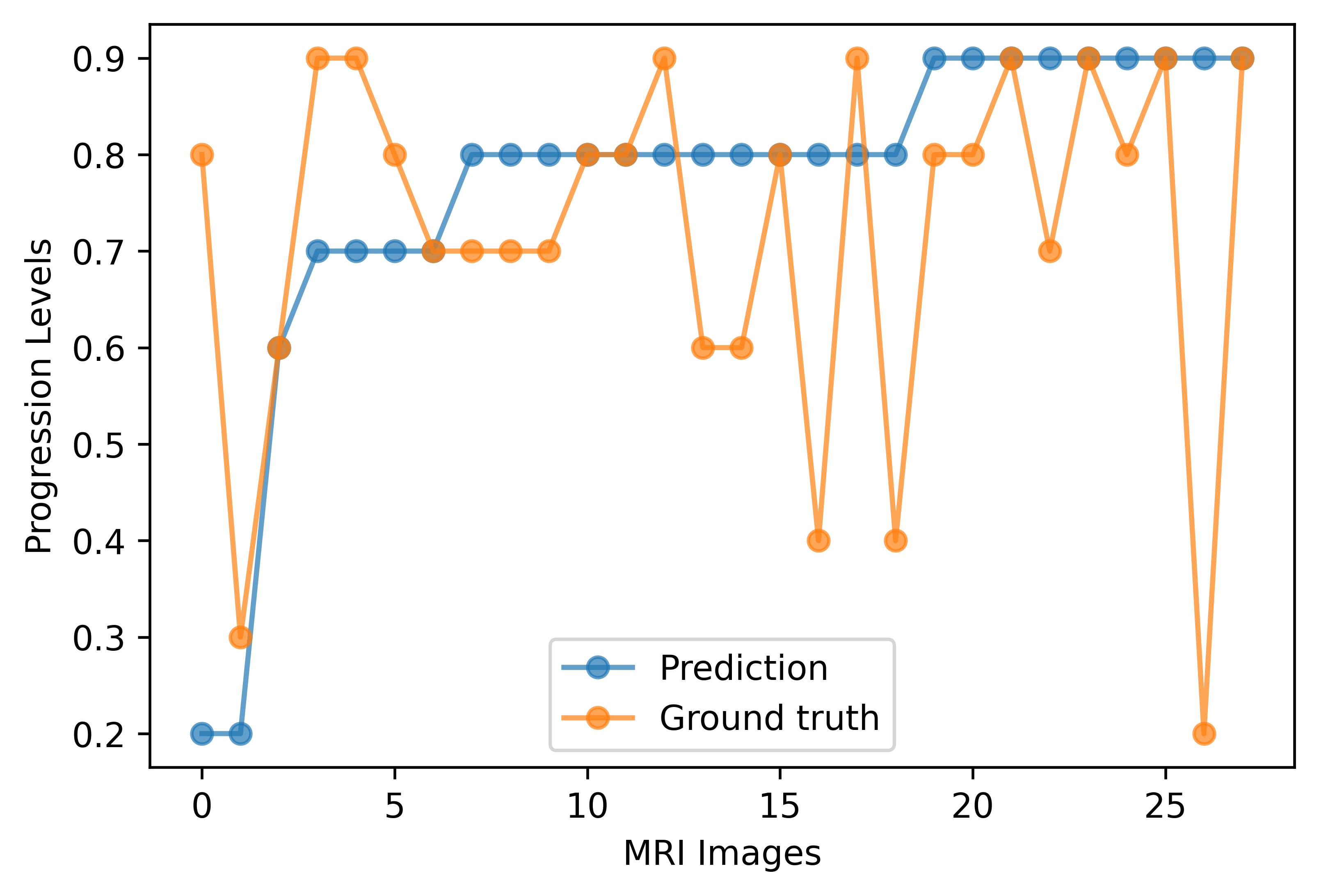} }}%
    \qquad
    \subfloat[\centering Unweighted Siamese]{{\includegraphics[width=5.6cm]{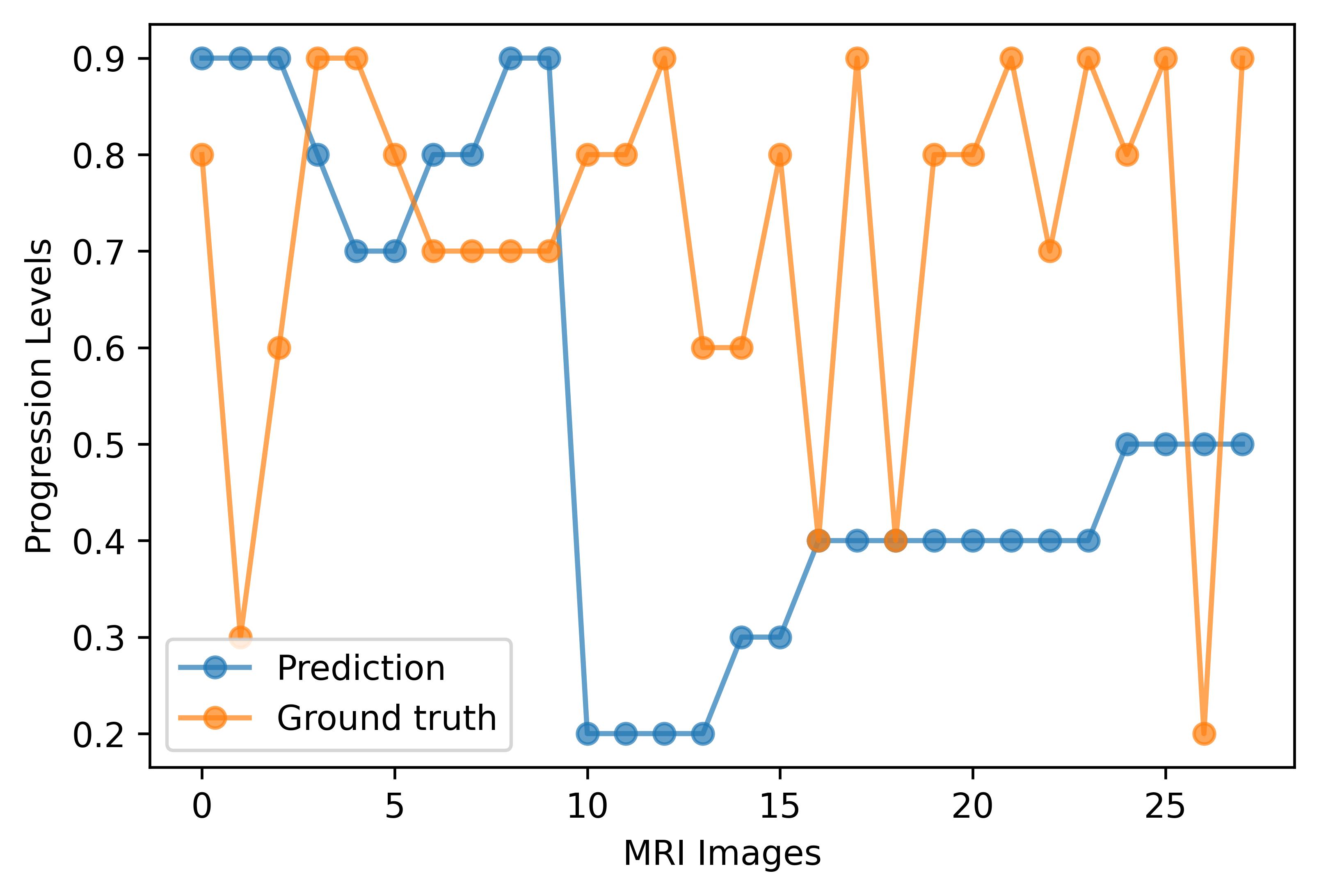} }}%
    
    \caption{Predicted progression levels of test MRI images against ground truth levels.}%
    \label{figure:predicted_progression_level}%
\end{figure}

A plot of predicted vs. ground truth MCI to AD progression levels is presented in Fig. \ref{figure:predicted_progression_level}. Our proposed Weighted Siamese network outperforms the Unweighted Siamese network at predicting progression levels(Fig. \ref{figure:predicted_progression_level} and Table \ref{table:average_mae_rmse}).

We observed that the simple modification of factoring the distance between an embedding of anchor and negative instances by a function of the progression level brought considerable performance gain in separating between the interpolated categories between MCI and AD. 

In Figure \ref{figure:predicted_progression_level}, although the Weighted Siamese outperforms the Unweighted Siamese, it also usually classifies the input test images with lower progression levels as if they are on a higher progression levels. This would mean brain images of patients that are far away from progressing to AD would be identified as if they are close to progressing. While it's important to correctly identify these low risk patients, we believe it's better to report the patients at lower risk as high risk and refer them for expert input than classifying high risk patients as low risk.

\begin{figure}[t]
    \centering
    \subfloat[\centering Training instances]{{\includegraphics[width=5.5cm]{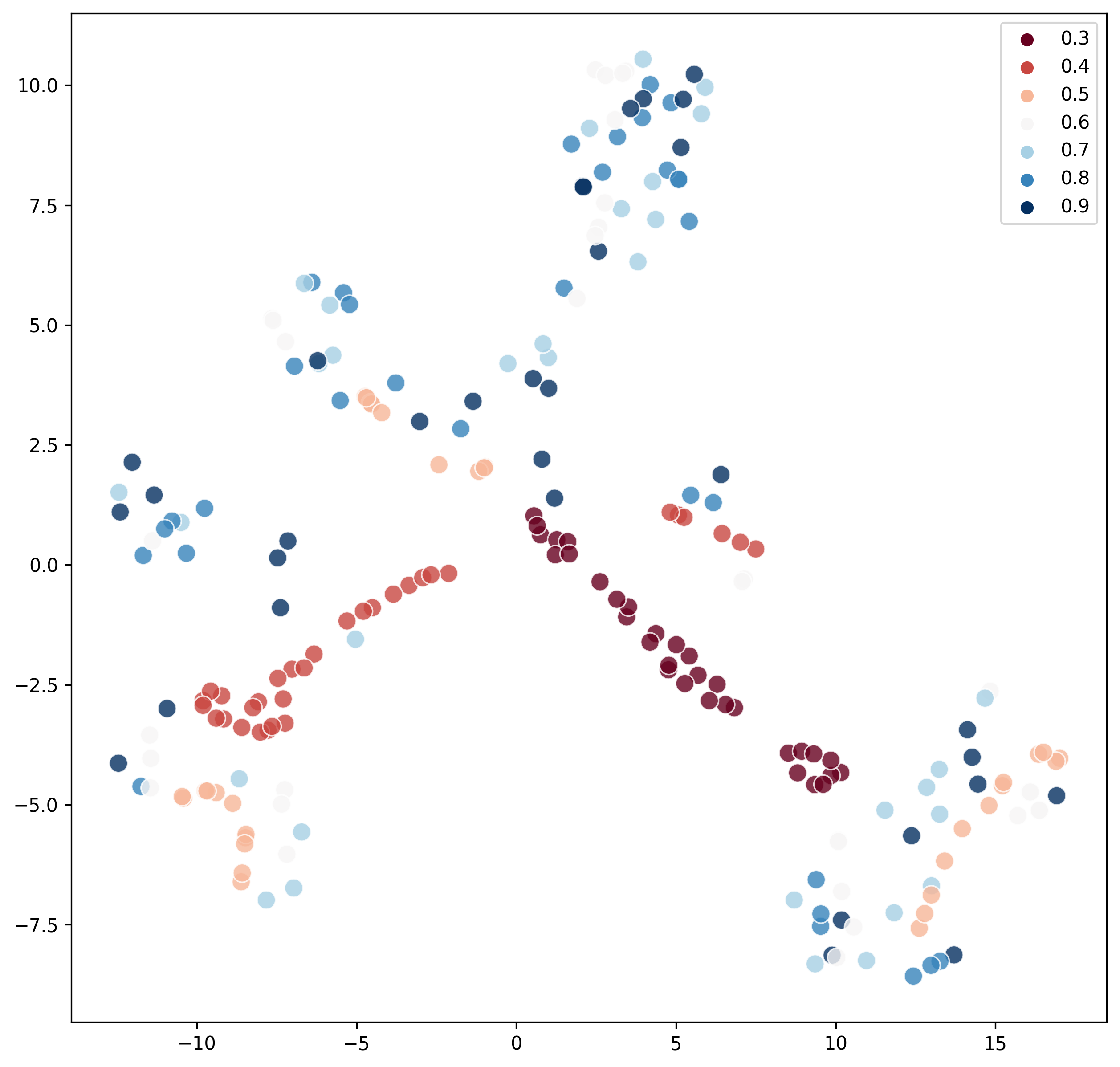} }}%
    \qquad
    \subfloat[\centering Test instances]{{\includegraphics[width=5.5cm]{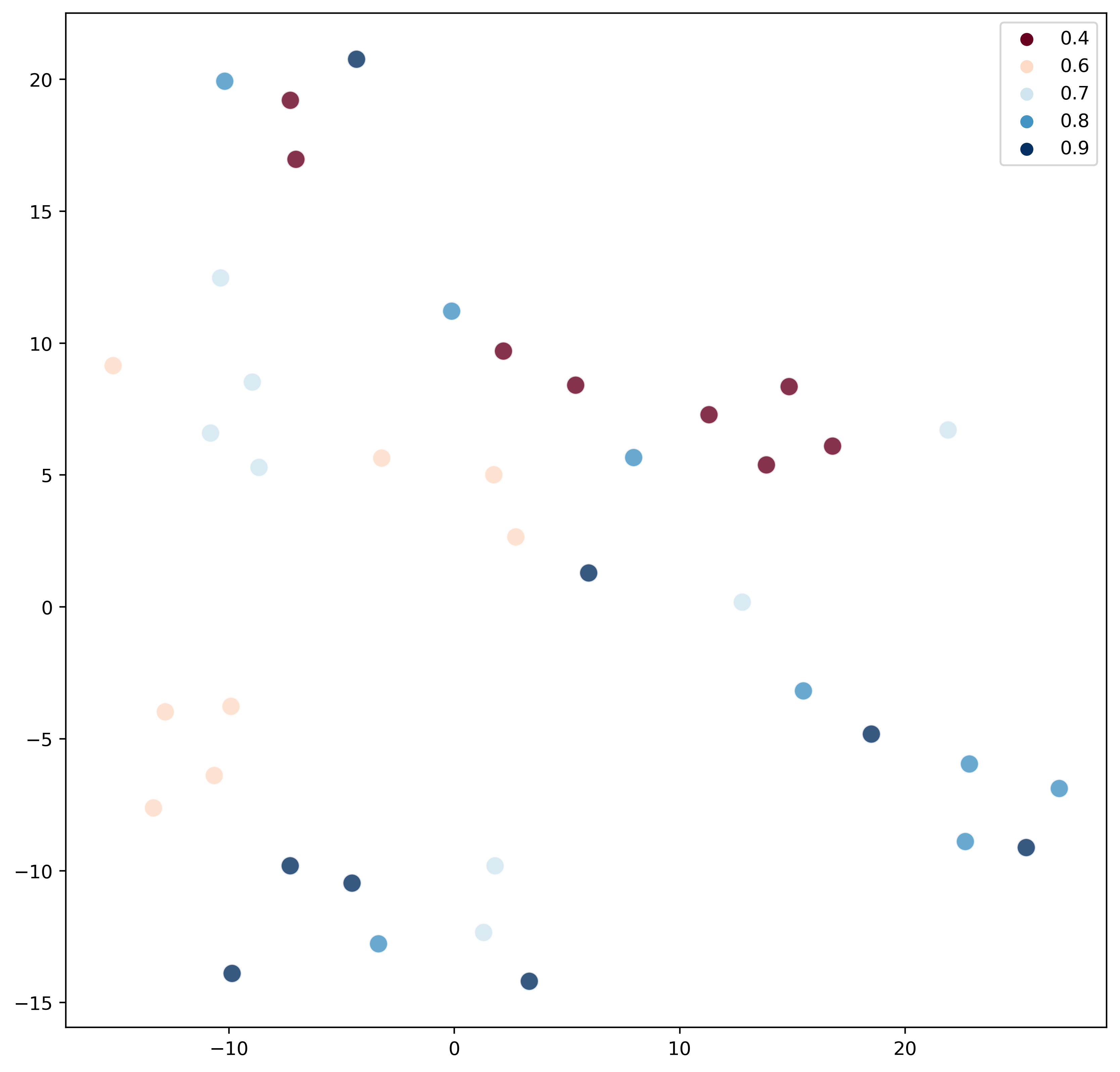} }}%
    
    \caption{Visualization of t-SNE of the embedding spaces of training and test instances.}%
    \label{figure:train_and_val_tsne}%
\end{figure}

An interpretation of the results of the proposed Weighted Siamese method using t-SNE is displayed in Figure \ref{figure:train_and_val_tsne}. The clustering of the embedding of input instances according to their progression levels, especially between the low-risk and high-risk progression levels assures us that the results represent the ground truth disease levels. 

\section{Conclusion}
\label{section:discussion}

Similarly to other image-based computer assisted diagnosis research work, the AD identification literature is heavily populated by disease stage classification. However, an interesting extra step can be taken to identify how far an input brain image is from progressing to a more severe stage of AD. We present a novel approach of interpolating ordinal categories in-between the MCI and AD categories to prepare a training dataset. In addition, we proposed and implemented a new Weighted loss term for Siamese networks that is tailored to such a dataset. With our experiments, we show that our proposed approach surpassed the performance of a model trained using a standard Unweighted loss term; and we show how the predicted levels translate to the ground truth progression levels by applying a model interpretability technique on the embedding space. We believe our approach could easily be transferred to other areas of medical image classification involving progressive diseases.


The diagnosis results taken in our study are bounded by the timeline of the ADNI study---meaning, even though based on extracted information a participant may have MCI during an examination year and they may progress to AD after some year(s), they could have had MCI before joining the ADNI study and their progression to AD might have taken longer than what we have noted. Future work should consider this limitation.


\section*{Acknowledgements}This publication has emanated from research conducted with the financial support of Science Foundation Ireland under Grant number 18/CRT/6183. For the purpose of Open Access, the author has applied a CC BY public copyright licence to any Author Accepted Manuscript version arising from this submission.

Data collection and sharing for this project was funded by the Alzheimer's Disease Neuroimaging Initiative (ADNI) (National Institutes of Health Grant U01 AG024904) and DOD ADNI (Department of Defense award number W81XWH-12-2-0012)\footnote{ADNI is funded by the National Institute on Aging, the National Institute of Biomedical Imaging and Bioengineering, and through generous contributions from the following: AbbVie, Alzheimer’s Association; Alzheimer’s Drug Discovery Foundation; Araclon Biotech; BioClinica, Inc.; Biogen; Bristol-Myers Squibb Company; CereSpir, Inc.; Cogstate; Eisai Inc.; Elan Pharmaceuticals, Inc.; Eli Lilly and Company; EuroImmun; F. Hoffmann-La Roche Ltd and its affiliated company Genentech, Inc.; Fujirebio; GE Healthcare; IXICO Ltd.; Janssen Alzheimer Immunotherapy Research \& Development, LLC.; Johnson \& Johnson Pharmaceutical Research \& Development LLC.; Lumosity; Lundbeck; Merck \& Co., Inc.; Meso Scale Diagnostics, LLC.; NeuroRx Research; Neurotrack Technologies; Novartis Pharmaceuticals Corporation; Pfizer Inc.; Piramal Imaging; Servier; Takeda Pharmaceutical Company; and Transition Therapeutics. The Canadian Institutes of Health Research is providing funds to support ADNI clinical sites in Canada. Private sector contributions are facilitated by the Foundation for the National Institutes of Health (www.fnih.org)  The grantee organization is the Northern California Institute for Research and Education, and the study is coordinated by the Alzheimer’s Therapeutic Research Institute at the University of Southern California. ADNI data are disseminated by the Laboratory for Neuro Imaging at the University of Southern California}.

\bibliographystyle{splncs04} 
\bibliography{splncs04}

\end{document}